\documentclass[%
prx,
superscriptaddress,
 amsmath,
 amssymb,
 twocolumn,
 aps,
 notitlepage,
]{revtex4-1}

\usepackage{graphicx}
\usepackage{dcolumn}
\usepackage{bm}
\usepackage{setspace}
\usepackage{xcolor}
\usepackage{textgreek}

\begin{document}

\title{Graphene's non-equilibrium fermions reveal Doppler-shifted magnetophonon resonances accompanied by Mach supersonic and Landau velocity effects.}

\affiliation{Department of Physics, Loughborough University, LE11 3TU, UK}

\affiliation{School of Physics and Astronomy, University of Nottingham, NG7 2RD, UK}

\affiliation{School of Physics and Astronomy, University of Manchester, Manchester, M13 9PL, UK}

\affiliation{National Graphene Institute, University of Manchester, Manchester, M13 9PL, UK}

\affiliation{Department of Physics, University of Lancaster, Lancaster, LA1 4YW, UK}

\affiliation{Tim Taylor Department of Chemical Engineering, Kansas State University, Manhattan, KS, 66506, USA}

\author{M.T. Greenaway}
\email{m.t.greenaway@lboro.ac.uk}
\affiliation{Department of Physics, Loughborough University, LE11 3TU, UK}
\affiliation{School of Physics and Astronomy, University of Nottingham, NG7 2RD, UK}

\author{P. Kumaravadivel}
\affiliation{School of Physics and Astronomy, University of Manchester, Manchester, M13 9PL, UK}
\affiliation{National Graphene Institute, University of Manchester, Manchester, M13 9PL, UK}

\author{J. Wengraf}
\affiliation{School of Physics and Astronomy, University of Manchester, Manchester, M13 9PL, UK}
\affiliation{Department of Physics, University of Lancaster, Lancaster, LA1 4YW, UK}

\author{L.A.~Ponomarenko}
\affiliation{School of Physics and Astronomy, University of Manchester, Manchester, M13 9PL, UK}
\affiliation{Department of Physics, University of Lancaster, Lancaster, LA1 4YW, UK}

\author{A.I.~Berdyugin}
\affiliation{School of Physics and Astronomy, University of Manchester, Manchester, M13 9PL, UK}

\author{J.~Li}
\affiliation{Tim Taylor Department of Chemical Engineering, Kansas State University, Manhattan, KS, 66506, USA}

\author{J.H.~Edgar}
\affiliation{Tim Taylor Department of Chemical Engineering, Kansas State University, Manhattan, KS, 66506, USA}

\author{R. Krishna Kumar}
\affiliation{School of Physics and Astronomy, University of Manchester, Manchester, M13 9PL, UK}

\author{A.K.~Geim}
\affiliation{School of Physics and Astronomy, University of Manchester, Manchester, M13 9PL, UK}
\affiliation{National Graphene Institute, University of Manchester, Manchester, M13 9PL, UK}

\author{L. Eaves}
\email{laurence.eaves@nottingham.ac.uk}
\affiliation{School of Physics and Astronomy, University of Nottingham, NG7 2RD, UK}
\affiliation{School of Physics and Astronomy, University of Manchester, Manchester, M13 9PL, UK}

\date{\today}

\begin{abstract}
Oscillatory magnetoresistance measurements on graphene have revealed a wealth of novel physics.  These phenomena are typically studied at low currents.  At high currents, electrons are driven far from equilibrium with the atomic lattice vibrations so that their kinetic energy can exceed the thermal energy of the phonons.  Here, we report three non-equilibrium phenomena in monolayer graphene at high currents: (i) a ``Doppler-like'' shift and splitting of the frequencies of the transverse acoustic (TA) phonons emitted when the electrons undergo {\it inter}-Landau level (LL) transitions; (ii) an {\it intra}-LL Mach effect with the emission of TA phonons when the electrons approach supersonic speed, and (iii) the onset of elastic inter-LL transitions at a critical carrier drift velocity, analogous to the superfluid Landau velocity. All three quantum phenomena can be unified in a single resonance equation.  They offer avenues for research on out-of-equilibrium phenomena in other two-dimensional fermion systems.
\end{abstract}

\maketitle

\section{Introduction}

Non-equilibrium phenomena in conventional semiconductors are of fundamental interest and have been exploited for technology, for example high performance GaAs Gunn oscillators \cite{Couch1989} and quantum cascade lasers \cite{Faist1994}. They are anticipated to be particularly prominent in graphene due to its weak electron-phonon coupling.

 The effect of high currents and strong electric fields, ${\bf F}$, on the magnetoresistance of Landau-quantised electrons in semiconductors has been investigated intensively.  In bulk n-type GaAs, fields $F\sim10^6$ Vm$^{-1}$ induce large shifts and splittings of the magnetophonon resonance (MPR) peaks due to scattering by 36 meV longitudinal optical phonons, along with quasi-elastic inter-Landau level (LL) transitions \cite{Eaves1984,Eaves1984a,Mori1988}.  In contrast, the field-induced shifts and splittings in high-mobility GaAs quantum well (QW) heterostructures arise from inter-LL scattering by acoustic phonons of lower energy (a few meV) \cite{Zudov2001,Dmitriev2012,Dmitriev2010,Zhang2008}.  A competing process due to elastic inter-LL interactions at high currents also give rise to Hall field-induced oscillations (HIRO) which have been investigated in QWs composed of GaAs \cite{Yang2002,Zhang2007,Zudov2017}, Ge/SiGe \cite{Shi2014}, and MgZnO/ZnO \cite{Shi2017}.  It is interesting to note that in GaAs QWs inter-LL processes can also manifest themselves as a breakdown of the integer quantum Hall effect \cite{Eaves1986,Tsui1983,Heinonen1984,Martin2003,Tomimatsu2020}.  

\begin{figure}[!t]
  \centering
\includegraphics[width=1\linewidth]{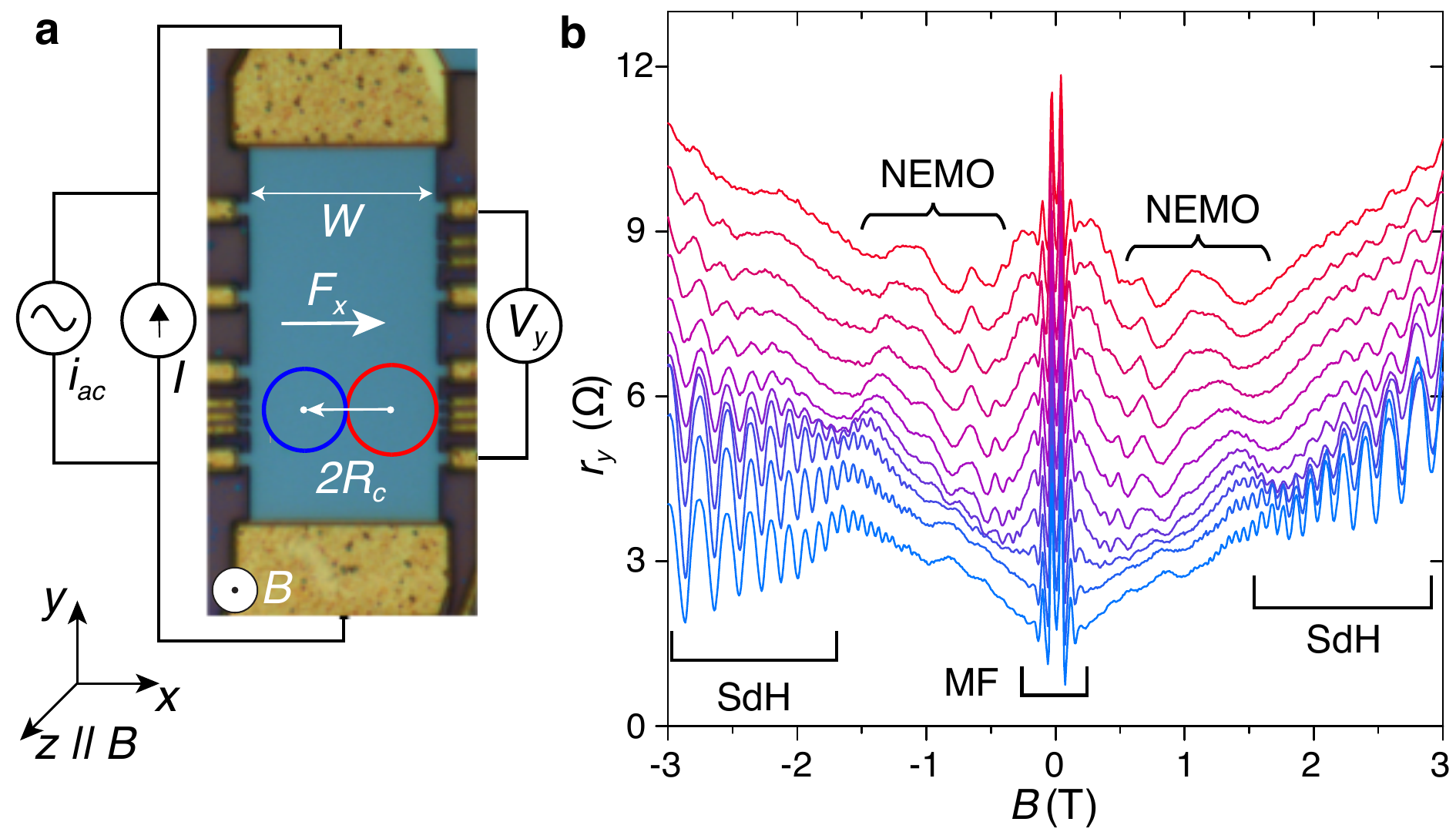}
  \caption{{\bf Current dependence of magnetoresistance oscillations in monolayer graphene Hall bars.} {\bf a} Optical micrograph image of the graphene Hall bar ($W=15$\textmu m) and a schematic diagram of the measurement configuration. {\bf b} Plots of differential resistance $r_{y}=dV_y/dI$ at $T=5$ K as a function of $B$ for DC currents, $I$, between 0 (blue) and 140 $\mu$A (red) in 14 \textmu A intervals, curves are offset by 0.7~$\Omega$ for clarity.  Curly brackets indicate the emergence of additional Non-equilibrium Magneto-Oscillations (NEMO) examined in detail in Figs \ref{fig:MPR} and \ref{fig:quills}.  Square brackets labelled SdH indicate Shubnikov-de Haas oscillations and square brackets labelled MF indicate the Magnetic Focusing peaks.  
\label{fig:schem}}
\end{figure}

For graphene, resonant magnetoresistance measurements under ``ohmic'' low current conditions have been used to study the quantum Hall and flux quantisation effects \cite{CastroNeto2009,Goerbig2011,Yankowitz2019,Novoselov2005,Zhang2005,KrishnaKumar2017}.  Non-equilibrium phenomena at higher currents are anticipated to be particularly prominent due to graphene's relatively weak electron-phonon coupling.  Furthermore the relativistic fermions in graphene have an order of magnitude higher group velocity than electrons in conventional semiconductors.  Similarly, graphene's acoustic and optical phonon modes have much higher speeds and larger energies. Graphene's ability to support large current densities of over $\sim100$ Am$^{-1}$ (or equivalently $\sim10^{11}$ Am$^{-2}$) enables us to explore resonant hot carrier phonon emission together with the formation and dissipation of magneto-excitons \cite{Kallin1984,Martin2003,Roldan2010,Girvin1986,Yang2018}.  We show that all of the observed phenomena can be unified in terms of a single generic resonant scattering equation.  

\section{Results and Discussion}

\subsection{Magnetophonon resonance splitting at high currents}
We search for non-equilibrium magneto-oscillations (NEMO) using large area Hall bars of high purity exfoliated monolayer graphene encapsulated between layers of exfoliated hexagonal boron nitride and mounted on a silicon oxide - silicon gate electrode, for further information regarding fabrication see Ref. \cite{Kumaravadivel2019}.  An image of the device is shown in Fig. \ref{fig:schem}a. Recent work has demonstrated that these devices, with dimensions well in excess of the phonon-limited mean free path ($\sim10$ \textmu m), are required to reveal magnetophonon resonances (MPR) in graphene \cite{Kumaravadivel2019,Greenaway2019}.  Those measurements were made at temperatures up to 200 K under ohmic conditions at small DC currents, so the carriers remained in equilibrium with the lattice phonons.   

Figure \ref{fig:schem}{a} shows our circuit arrangement.  In parallel with a small low frequency AC modulation current, $i=2$ \textmu A, a DC current, $I$, up to $I=1$ mA is passed through a multi-terminal Hall bar of width $W=15$ \textmu m.  The differential magnetoresistance $r_{y}=dV_{y}/dI$ is measured using a lock-in amplifier as a function of $I$ and applied magnetic field, $B$.

\begin{figure*}[!t]
  \centering
\includegraphics[width=1\linewidth]{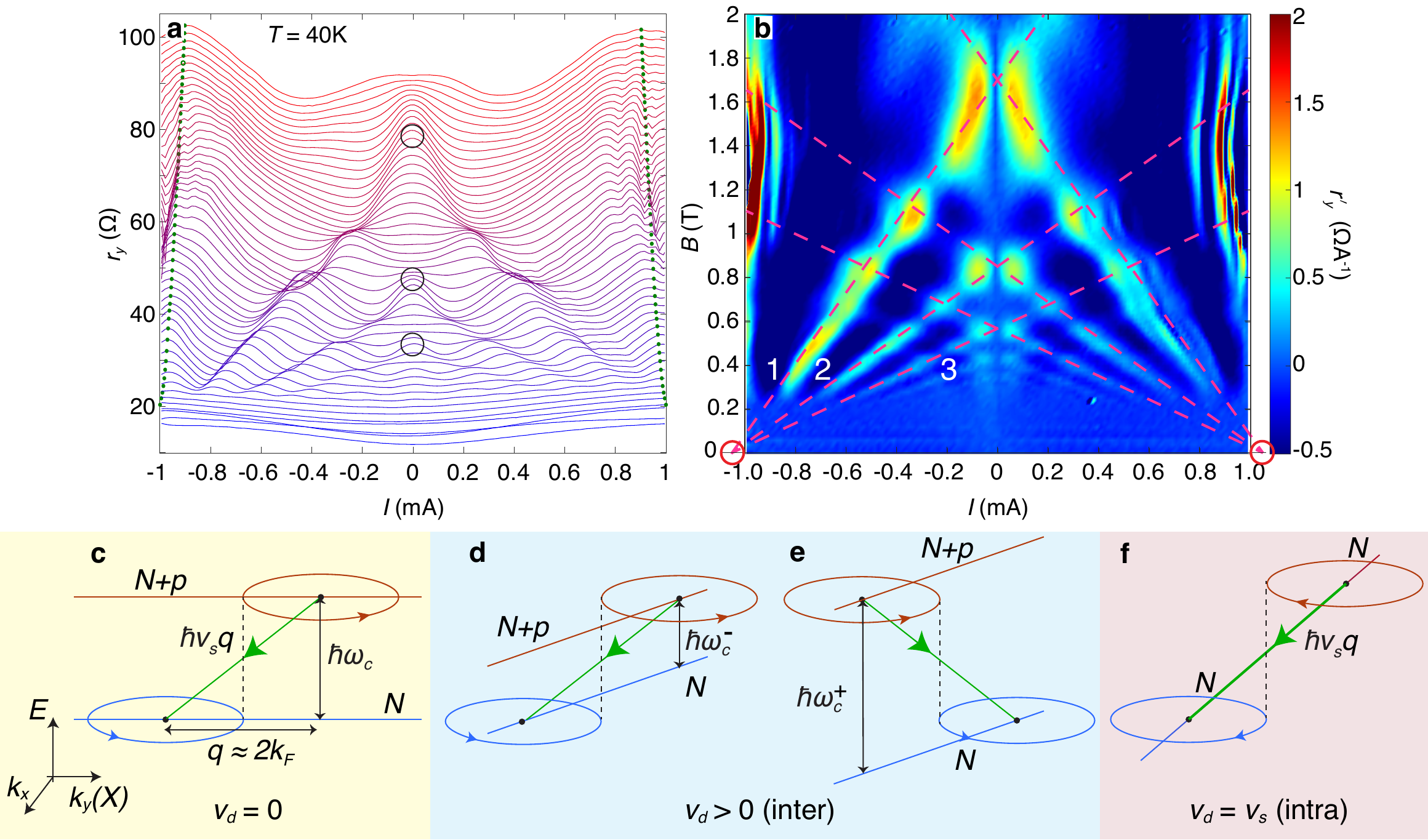}
  \caption{{\bf Non-equilibrium magnetoresistance oscillations at $T=40$ K: magnetophonon resonance splitting and the Mach effect.} {\bf a} Plot of differential resistance $r_{y}(I)$ for $B$ between 0 to 2 T in 0.04 T intervals and {\bf b} the corresponding colour map of $r_{y}'(I,B)=-{\rm sgn}(I)dr_{y}/dI$, measured at $T=40$ K and $n=3.16\times10^{12}$ cm$^{-2}$.  In {\bf a} the curves are offset by 1.5 $\Omega$ and the open circles indicate the magnetophonon resonance peaks $p=1,2,3$ under ohmic conditions when $I=0$.  The green markers show the position of the peak in the resonance at $v_d\approx v_s$ given by Supplementary Eq. (1).  In {\bf b} the dashed magenta lines are a fit given by Eq. (\ref{eq:Bpvd}) when $v_s=v_{TA}$ and $p=1,2,3$ (labelled).  The red circles at the left and right hand corners of the colour map highlight the point of convergence of the magenta lines when $B=0$.  Figures {\bf c-f} show energy-momentum diagrams in applied $B$ and $F_x$ to demonstrate the semiclassical conditions for strong scattering between initial (red, $N+p$) and final (blue, $N$) Landau states when: {\bf c}, $v_d\approx0$, corresponding to ohmic MPR; {\bf d} and {\bf e} when $v_d<v_s$, show Hall field-induced splitting of the MPR features when $B=B_p^-$ ({\bf d}) and $B=B_p^+$ ({\bf e}). Panel {\bf f}: $v_d=v_s$, corresponds to intra-LL acoustic phonon scattering.  In panels {\bf c}-{\bf f}: the red and blue circles show the semiclassical trajectories of the carriers in $k-$space, the red and blue lines show the energies of the initial and final Landau states and their dependence on $k_y(X)$ in the finite Hall field. The dashed vertical line shows the points of intersection of the semiclassical figure-of-8 orbits when $q=2k_F$. The arrowed green lines show the phonon-assisted transitions between LLs, panels {\bf c}-{\bf e}, and within a single LL, panel {\bf f}. 
\label{fig:MPR}}
\end{figure*}

\begin{figure*}[!t]
  \centering
\includegraphics[width=1\linewidth]{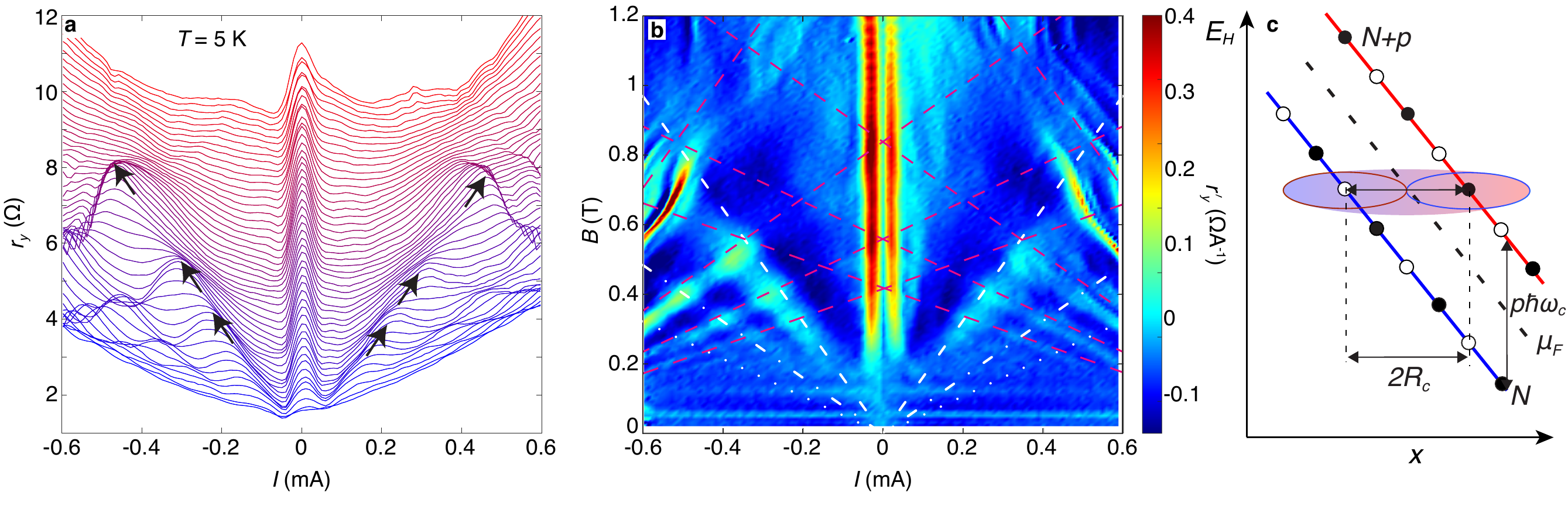}
  \caption{{\bf Non-equilibrium magnetoresistance oscillations at $T=5$ K: elastic inter-LL transitions.} {\bf a} Plot of $r_{y}(I)$ for values of $B$ between 0 to 1.2 T in 0.04 T intervals when $T=5$ K and $n=3.16\times10^{12}$ cm$^{-2}$.  Curves are offset by 1.5 $\Omega$ for clarity. Black arrows highlight position of peaks in $r_y$  {\bf b} Colourmap of $r'_{y}(I,B)$ when $T=5$ K.  Dashed white lines show the linear dependence of the $B_p$ on $I$, see Eq. \ref{eq:Bpquills}, for the phononless resonances $p=1,2$ and $3$.   In {\bf b} the magenta lines are a fit given by Eq. (\ref{eq:Bpvd}) for the Hall-field dependent MPR peak with $p=1,2,3$ and 4.  Panel {\bf c} shows how elastic inter-LL transitions can create a non-equilibrium distribution of carriers involving interacting electron-hole pairs, ``magneto-excitons''.  
\label{fig:quills}}
\end{figure*}

Figure \ref{fig:schem}b shows the dependence of $r_{y}$ on $B$ and $I$ at a bath temperature of $T=5$ K and a gate voltage $V_g=-60$ V which generates a carrier sheet density (holes) of $n=3.16\times10^{12}$ cm$^{-2}$.  The large amplitude features close to $B=0$ which persist up to currents of at least $\sim 250$ \textmu A and over a wide range of $n$ (see Supplementary Figure 1) are due to well-established magnetic focusing (MF) of the quasi-ballistic carriers \cite{Taychatanapat2013, Kumaravadivel2019}. They indicate that our carefully exfoliated devices are largely free from disordered localised states.  At low currents, strong Shubnikov-de Haas (SdH) oscillations appear at $B\gtrsim 1$ T.  They damp out as $I$ is increased and the carrier distribution becomes non-thermal and broadened around the Fermi level, $\mu_F$. In their place, a new set of resonant magneto-oscillations appear (NEMO, see Fig. \ref{fig:schem}b) whose behaviour and origin we explore in Figs. \ref{fig:MPR} and \ref{fig:quills}.  

The effect on the magnetoresistance oscillations of the Hall field, $F_x$, induced by the DC current is illustrated in Figs. \ref{fig:MPR}a and b which plot a set of $r_{y}(I)$ curves versus $B$.  At $T=40$ K and $I=0$, the peaks in $r_{y}$ (circled), which are periodic in $1/B$, correspond to ohmic MPR oscillations due to TA phonon scattering transitions, as reported recently \cite{Kumaravadivel2019,Greenaway2019} and illustrated schematically in Fig. \ref{fig:MPR}c.  With increasing $I$ and $F_x$, the magnetic field positions of the magnetophonon resonances undergo large shifts and splittings.  These features are highlighted by magenta dashed lines in Fig. \ref{fig:MPR}b which plots the second derivative of $V_{y}$ with respect to $I$,  $r'_{y}=-{\rm sgn} (I)dr_{y}/dI$. 

The splittings and shifts arise from the effect of $F_x$ which tilts the energies of the quantised Landau states, see Figs. \ref{fig:MPR}d-f.  Graphene's high carrier mobility, $\mu\sim 50$ m$^2$V$^{-1}$s$^{-1}$, provides a large value of $\mu B\gg1$, thus ensuring Landau quantisation of well-defined cyclotron orbits at fields as low as $B\sim0.3$ T.  When $F_x=0$, a carrier in a LL with index $N$ has an energy $E_N=\hbar\sqrt{2 N} v_F/ l_B$, where $l_B=\sqrt{\hbar/eB}$ is the magnetic length and $v_F$ is the Fermi velocity in graphene.  When $F_x\neq0$ the energy of a carrier depends on the location of its real-space orbit centre $X$ across the width of the Hall bar and thus on its wave-vector component, $k_y$, along the $y$ axis: $X(k_y)=-k_yl_B^2$, in the Landau gauge. Therefore, under the condition $v_d\ll v_F$, which is satisfied in our experiment, the energy of a carrier in LL index $N$ is given by \cite{Lukose2007,Raichev2020}:
 \begin{equation}
E_N(k_y)=E_N - e F_x X(k_y).
\label{eq:refquantQcond}
\end{equation}

A carrier can scatter inelastically between LL states with indices $N$ and $(N+p)$ and wavevectors $k_y$ and $k_y'$  by emitting or absorbing an acoustic phonon of speed $v_s$ and wavevector $q$ governed by the relation
 \begin{equation}
E_{N+p}(k'_y)-E_N(k_y)=\pm\hbar v_s q.
\end{equation}
To a good approximation, the energy gap between LLs is given by the correspondence principle when $N$ is large and $p$ is small:
\begin{equation}
E_{N+p}-E_N=p\hbar\omega_c.
\label{eq:energywc}
\end{equation}
Here, $p=1,2,3...$, $\omega_c=eB/m_c$ is the cyclotron frequency and $m_c=\hbar k_F/v_F$ is the wave-vector dependent cyclotron mass for Landau states close to the Fermi energy, where $N\approx 30$ at $B=1$ T.  

Scattering between Landau states is strongest when the spatial overlap of their wavefunctions and the transfer of momentum is at a maximum. Semiclassically, this occurs when the two cyclotron orbits of the initial and final states just touch to form a ``figure-of-8'' scattering configuration, see Fig. \ref{fig:schem}a and Figs 2c-f \cite{Greenaway2015,Greenaway2019}. The separation between the two orbit centres is given by $2R_c$, where $R_c=l_B^2 k_F$ is the classical cyclotron radius at the Fermi energy and $k_F=\sqrt{\pi n}$ is the Fermi wavevector.  For such a transition, the carrier's momentum is changed by $2 \hbar k_F$ along the $x$-axis.

\subsection{Doppler-shifted frequencies of emitted phonons}

Referring to Figs 2d and e, and using equations (\ref{eq:refquantQcond})-(\ref{eq:energywc}) with the phonon-wavevector, $q=2k_F$ we derive a generic condition for current-dependent magnetophonon scattering between LLs:
\begin{equation}
p\omega_c =2 k_F (v_s\pm v_d),
\label{eq:genres}
\end{equation}
where $v_d=F_x/B$ is the carrier drift velocity along the Hall bar. To a good approximation
\begin{equation}
v_d=\frac{I}{enW},
\label{eq:drift}
\end{equation}
since $\mu B\gg1$.  In the limit when $I=v_d=F_x=0$, we obtain the ohmic MPR condition \cite{Kumaravadivel2019,Greenaway2019}, $p\omega_c=2v_sk_F$ or equivalently
\begin{equation}
B_p(v_d=0)=B_p^0=\frac{hnv_s}{pev_F}.
\label{eq:MPR}
\end{equation}
The measured values of $B^0_p$ and their dependence on $n$ give the value of $v_s$ and confirm that the resonances are due to TA phonon scattering with $v_s=v_{TA}=1.36\times10^4$~ms$^{-1}$   \cite{Kumaravadivel2019}.  Eq. \ref{eq:genres} can be rewritten to give the magnetic field position of each resonance in terms of $I$:
\begin{equation}
B_p(I)=B_p^0\left(1 \pm \frac{I}{v_senW}\right).
\label{eq:Bpvd}
\end{equation}
Thus for a given $I$ and $p$ there are two resonant conditions with magnetic field positions $B_p(I)^+$ and $B_p(I)^-$.  They correspond to scattering of carriers either in the positive or negative $x-$direction for which the electrostatic potential energy,  $eF_x x$, increases or decreases.  The shifts of the phonon frequencies, $\omega_s$, are found by rearranging Eq. \ref{eq:genres} to give $\omega_s=2k_Fv_s=\omega_c\pm2k_Fv_d$, see also Figs \ref{fig:MPR}d and e.  These shifts can be thought of as a modified form of the Doppler effect,   in which phonon quanta of the TA sound waves are preferentially emitted perpendicular to the Hall field.

As shown in Fig. \ref{fig:MPR}b the dashed magenta lines corresponding to relation (\ref{eq:Bpvd}) provide a good fit to the loci and splitting of the measured magnetophonon resonances in $r'_{y}$. Over this range of $I$ and $B$ with $n=3.16\times10^{12}$ cm$^{-2}$, the inelastic resonant transitions between partially filled LLs occur around $\mu_F=210$ meV, with large LL indices ($N\sim 33$ at $B=1$ T) and with approximately equal LL spacing  $\hbar\omega_c=ev_FB/k_F\sim3$ meV.  A notable feature in Fig. \ref{fig:MPR}b is that the peaks in $r'_{y}$ are strongest at the crossing points of the magenta lines when $B_{p'}^+=B_p^-$ so that two different energy relaxation routes are available.  

\subsection{Supersonic electrons and a Mach effect}

We consider next the strong and broad peak in $r_{y}$ in Figs. \ref{fig:MPR}a centred at $I\approx \pm 0.9$ mA over a wide range of $B$ from $\sim0.5$ to 2 T.  The position of the peak is marked by the near-vertical set of green dots in Fig. \ref{fig:MPR}a, see Supplementary Note 1. Beyond this value of $|I|$, $v_d\geq1.19\times10^{4}$ ms$^{-1}$ and approaches the measured value of $v_{TA}=1.36\times10^4$ ms$^{-1}$ in this device \cite{Kumaravadivel2019,Greenaway2019}.  This leads to the onset of strong {\it intra}-LL scattering of fast-drifting carriers and the emission of TA phonons, including those with large momentum and wavevector $q\approx2k_F$ (see Fig.~\ref{fig:MPR}f) which make a major contribution to the magnetoresistance.   For such an intra-LL transition, the energy conservation condition for phonon emission is independent of the energy separation, $\hbar\omega_c$, of the LLs. This results in the broad, large amplitude peak in $r_y$ at $I\approx \pm 0.9$ mA with increasing $B$ from 0.5 to 2 T. This corresponds to the generic resonance condition, Eq. \ref{eq:genres} with $p=0$ so that 
\begin{equation}
\frac{I}{enW}=v_d=v_s
\label{eq:super}
\end{equation}
is independent of $B$. This strong peak over such a wide range of $B$ in monolayer graphene represents a  2D analogy with Mach's ``supersonic boom'' effect \cite{Dmitriev2010,Zhang2008,Heinonen1984,Esaki1962}.  In this regime, the electric field effectively acts as an energy pump for amplified phonon emission by the hot carriers \cite{Hutson1961,Spector1963,Eckstein1963}. Note that as $|v_d|\rightarrow v_s$  then $B_p(I)^-\rightarrow0$, see Eq. \ref{eq:Bpvd}.  This is confirmed by the convergence of the shifted magnetophonon resonance peaks in $r_y'$ at the points $I=\pm1.05$ mA and $B=0$, red circles in Fig. \ref{fig:MPR}b. The magenta lines in Fig. \ref{fig:MPR}b highlight this convergence. The linewidth, $\Gamma(B)\propto\sqrt{B}$, of the peak has a weak $B$-dependence similar to previous experiments and theoretical work on graphene \cite{Funk2015}, see Supplementary Note 1.  

\subsection{Elastic inter-LL transitions}

At low temperatures ($T=5$ K), we observe a third type of non-equilibrium resonant phenomenon which is revealed by the well-defined resonant peaks in $r_y$ highlighted by black arrows in Fig. \ref{fig:quills}a.  The dashed white lines in the $r'_{y}$ plot in Fig. \ref{fig:quills}b indicate the strong peaks which are periodic in $1/B$ and shift linearly with a slope $dB/dI\approx\pm1.6$ T mA$^{-1}$.  Two weaker peaks (dot-dashed lines) shift at rates of 0.8 T mA$^{-1}$ and $\approx 0.5$ T mA$^{-1}$. These peaks have a quite different character from the shifts of the MPR peaks observed at 40 K in Fig. \ref{fig:MPR}.  We now demonstrate that this third type of resonance arises from energy-conserving transitions between LLs, analogous to Zener tunnelling \cite{Zener1934}, as shown schematically in Fig. \ref{fig:quills}c.  They correspond to the semiclassical ``figure-of-8'' condition when $p\hbar\omega_c=2k_Fv_d=2R_ceF_x$ (i.e. generic equation Eq. \ref{eq:genres} with $v_s=0$) from which we obtain a critical carrier drift velocity
\begin{equation}
v_c=p\frac{\omega_c}{2k_F}.
\label{eq:vdcond}
\end{equation}
The resonances occur at magnetic field $B_p$ which shifts linearly with current: 
\begin{equation}
B_p=\frac{hnv_d}{pev_F}=\frac{h}{p e^2 v_F W} |I|.
\label{eq:Bpquills}
\end{equation}

The white lines in Fig. \ref{fig:quills}b given by Eq. \ref{eq:Bpquills} with $p=1$, 2 and 3 and~$v_F=1.05\times10^6$ ms$^{-1}$ provide a good fit to the measured linear $B(I)$ dependence of the resonant peaks in $r'_y$.   Note that Eq. \ref{eq:Bpquills} is independent of $n$ (see also Supplementary Note 2) and involves only one fitting parameter, $v_F$; therefore, our measurements provide an independent method to determine the Fermi velocity in graphene.  This measurement of $v_F$ is arguably more direct than that obtained by modelling the temperature-dependent damping of SdH oscillations \cite{Novoselov2005}.  We suggest that non-equilibrium magneto-oscillation measurements combined with Eq. \ref{eq:vdcond} could be useful to determine the cyclotron effective mass of other 2D materials. 

\subsection{Magneto-excitons}

These elastic phononless inter-LL transitions require a mechanism, such as defect-induced scattering \cite{Vavilov2007,Raichev2020,Eaves1984}, that enables transitions between the orthogonal initial and final states. In our encapsulated graphene Hall bars, the high carrier mobility, long mean free path \cite{Kumaravadivel2019} and the presence of magneto-focusing peaks in the magnetoresistance, labelled MF in Fig. \ref{fig:schem}b, suggest that defect-induced carrier scattering is relatively weak. Therefore, we propose a particular type of electron-electron interaction, namely the formation of magneto-excitons \cite{Yang2018,Kallin1984,Martin2003,Roldan2010,Girvin1986}, as a likely contributor to the strong resonant processes described by Eq. \ref{eq:Bpquills}. Magneto-excitons are quasi-particles which comprise an interacting electron and hole pair in energetically adjacent LLs .  They are connected with the poles in the polarizability function, $\Pi(\omega,q)$, at $q=0$ and $\omega=p\omega_c$ $(p=1,2...)$ \cite{Kallin1984}.  In the absence of an electric field, the magneto-exciton dispersion is gapped so that $\omega^p_{ME}(q,v_d=0)>p\omega_c$, and there is a low probability for exciton formation.   However, in a strong Hall field, the magneto-exciton dispersion is modified by the term $v_d q$: 
\begin{equation}
\omega^p_{ME}(q,v_d)=\omega^p_{ME}(q,v_d=0)-v_d q.
\end{equation}
When $q> \omega^p_{ME}(q,v_d=0)/v_d\approx \omega_c/v_d$, $\omega^p_{ME}(q,v_d)$ is negative, the magneto-exciton gap closes \cite{Yang2018}. 

A magneto-exciton with a particular $q$ can only form when its spectral density is non-zero which, within a single particle picture, corresponds to a strong overlap of the electron and hole wavefunctions of the $N$ and $(N+p)$ LL states \cite{Yang2018}.  The onset of this condition corresponds to the ``figure-of-8'' configuration shown in Fig. \ref{fig:quills}c in which magneto-excitons form with a length scale $\approx2R_N$ and wavevector $q\approx 2k_F$.  Thus, when $v_d\gtrsim \omega_c/2k_F=v_c$ spontaneous formation and proliferation of magneto-excitons can occur.  Strong evidence for such a process has been detected as shot noise in quantum Hall effect breakdown in bilayer graphene \cite{Yang2018}.  A magneto-exciton can decay by dissociation in a Hall field or by scattering of the component charges into nearby LL states with lower energy, accompanied by the emission of phonons, thus generating the dissipative voltage $V_y$ which gives rise to the resonant magnetoresistance peaks in Fig. \ref{fig:quills}a and b.  The enhancement of the resonances in $r_{y}$ and $r'_{y}$, revealed in Fig. \ref{fig:quills}b at the intersections of the conditions for Hall-field shifted magnetophonon resonance (magenta lines) and magneto-exciton formation (dashed white lines), can then be explained by the combined effect of magneto-exciton formation and decay by phonon emission.

Finally, we note that the formation of magneto-excitons when $v_d\geq v_c=p\omega_c/2k_F$, see Eq. \ref{eq:vdcond}, is somewhat analogous to the formation of quasi-particles and dissipation in superfluids \cite{Landau1941,Girvin1986} and that the expression for our critical velocity, $v_c$, resembles the critical Landau velocity, given by the energy-momentum ratio of a superfluid quasi-particle. For Landau-quantised carriers in graphene, the circulation, $\kappa$, around a cyclotron orbit is given by:
\begin{equation}
\kappa_N=\oint {\bf v}.d{\bf l} = 2\pi R_c v_F=\frac{2\pi}{eB} E_N.
\end{equation}
When a magneto-exciton forms due to an inter-LL transition with $N\rightarrow N+1$ ($p=1$), the change in circulation $\Delta\kappa=\kappa_{N+1}-\kappa_N=2\pi\hbar\omega_c/eB=h/m_c$ is ``quantised''.  In superfluids, the circulation quantum can be envisaged as a ``quantum'' of viscosity \cite{Lvov2014,Trachenko2020}.  For a sheet density of $n=3.2\times10^{12}$ cm$^{-2}$, $h/m_c\approx0.02$ m$^2$s$^{-1}$.  Therefore, at the onset of dissipation when $v_d=v_c$, the dimensionless ratio ${\rm R}=v_c W/\Delta \kappa= eBW/2hk_F(\approx6$ at $B=1$ T), acts rather like a critical Reynolds number in classical hydrodynamics.  

In conclusion, we have measured the oscillatory magnetoresistance of large area, high mobility monolayer graphene Hall bars over a range of currents up to 1 mA. They reveal three distinct phenomena involving Hall field-dependent magnetophonon resonance, behaviour analogous to Mach supersonics at intermediate temperatures (40 K) and a critical Landau velocity at low temperatures (5 K).  All three quantum phenomena can be described by a semiclassical generic equation for intra- and inter-LL transitions.

\section*{Acknowledgements}
This work was supported by the Engineering and Physical Sciences Research Council (grant numbers EP/V008110/1 and EP/T034351/1), the ERC (VANDER) and the Lloyd's Register Foundation.  Support for hBN crystal growth from the Office of Naval Research (award no. N00014-20-1-2474) is appreciated.

\newpage

\newpage\hbox{}\thispagestyle{empty}\newpage

\section*{Supplementary Note 1:  Broadening of peak in $r_y$ due to {\it intra}-LL scattering}

The width of the broad peak in $r_{y}$,  marked by green dots in Fig. 2a of the main text, and its amplitude both increase steadily with increasing $B$.  This is consistent with the energy broadening of LL width, $\Gamma\sim\sqrt{B}$, observed in previous experiments and theoretical work [Ref. 40 of the main text].  Intra-LL scattering gives rise to an increase in the dissipative voltage, $V_y$ with a maximum at $v_d=v_s$.  Assuming the peak in $V_y$ has a Gaussian form, as it broadens the peak in the differential resistance, $r_y=dV_y/dI$, will be shifted by a small amount 
\begin{equation}
v_d=v_s\pm\frac{\Gamma(B)}{2k_F\hbar}
\label{eq:broad}
\end{equation}
for negative and positive values of the current, $I$, The green dots in Fig. 2a of the main text show that Eq. \ref{eq:broad} provides a good fit to the position of the broad peak in $r_y$ for $B\approx 0.5$ T and $B\approx 2$ T when $\Gamma=\gamma\sqrt{B}$ with $\gamma=0.5$~meV~T$^{-1/2}$,  which is in good agreement with the LL broadening parameters extracted from the full quantum simulations of MPR, as discussed in Ref. [31] of the main text.  

\begin{figure}[!t]
  \centering
\includegraphics[width=0.6\linewidth]{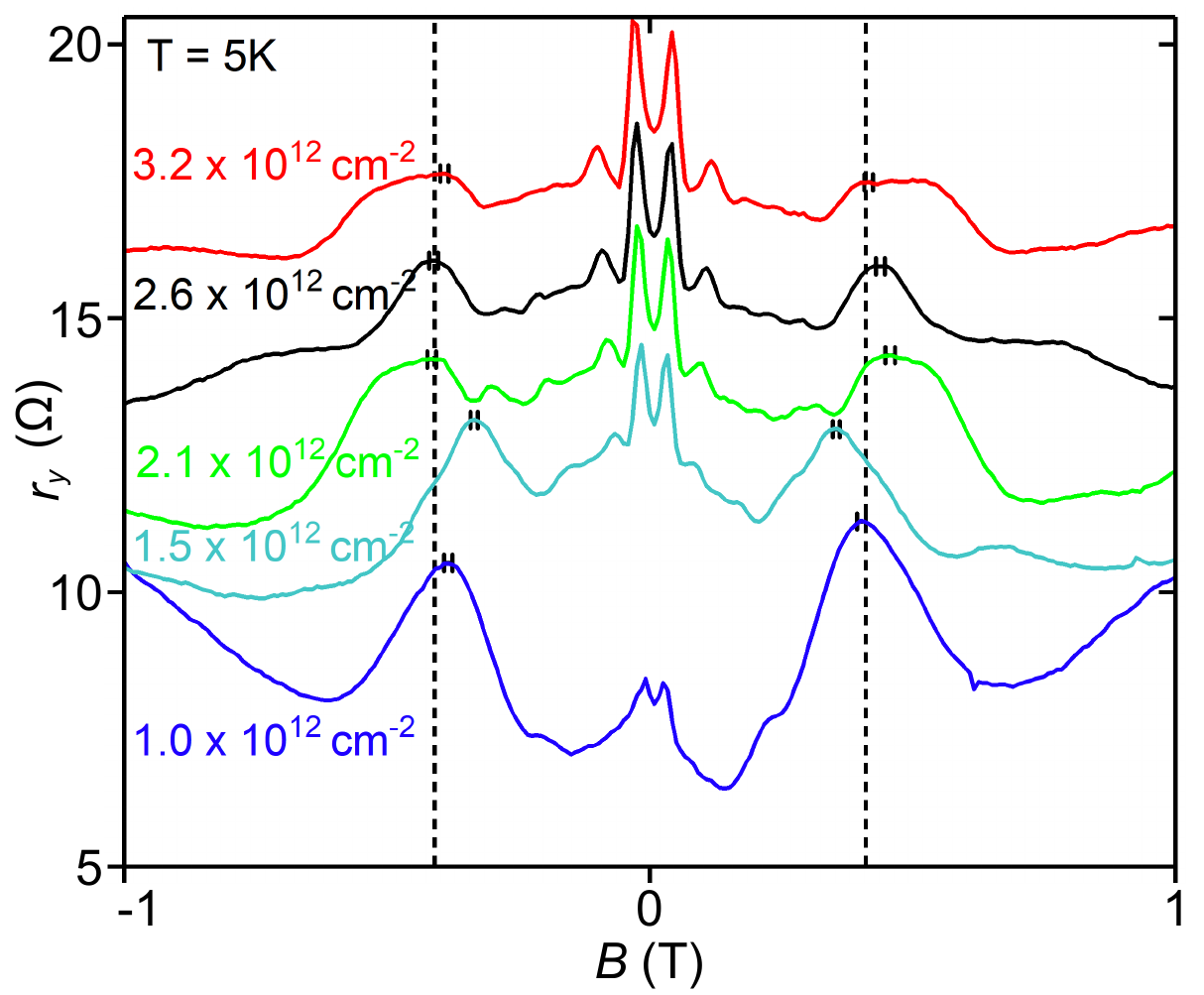}
  \caption{Supplementary Figure 1 {\bf Carrier density dependence of magnetoresistance oscillations} when $T=5$ K and $I=-250$ \textmu A for hole carrier densities ranging from $n=1\times10^{12}$ cm$^{-2}$ to $n=3.2\times10^{12}$ cm$^{-2}$, labelled in the figure.  Vertical dotted lines highlight the magnetic field-independent position of the resonance condition for elastic inter-LL  transitions given by Eq. (10) of the main text.  The magnetic field corresponding to the maximum in $r_y$ is indicated by the small ``H'' marker.   The oscillations around $B=0$ arise from resonant magnetic focusing.  
\label{fig:carrden}} 
\end{figure}

\section*{Supplementary Note 2: Carrier density dependence of magneto-oscillations at low temperatures}

Supplementary Fig. \ref{fig:carrden} shows the carrier density dependence of $r_y(B)$ at low temperature,  $T=5$ K.   At low magnetic field, around $\pm0.1$ T, we observe a series of sharp peaks corresponding to the magnetic focusing (MF) of ballistic electron trajectories as discussed in the main text.  In addition, each curve reveals a strong and broad maximum in the differential magnetoresistance at $B\sim 0.4$ T.  This value corresponds closely with the elastic inter-LL transition resonance condition of $B=0.41$ T given by Eq. (10) of the main text,  see dotted vertical lines in Fig. \ref{fig:carrden}. We suggest that the small differences in the magnetic field position of these peaks is due to the competing interaction of the elastic transitions and resonant magnetophonon emission, see discussion in main text.  We find that the strength of the peak in $r_y(B)$ at $B=0.41$ T decreases with increasing carrier density, due the increased separation of the LLs at low carrier densities.  In our manuscript we focus on detailed measurements with a carrier density of $n=3.2\times10^{12}$ cm$^{-2}$ as we find this is the optimum $n$-value to reveal the interaction of the elastic process with the splitting and shifts of the MPR peaks, which become more defined at high carrier density, see Ref. [30] of the main text.


\begin{thebibliography}{10}
 
\bibitem{Couch1989} Couch, N.~R., Spooner, H., Beton, P.~H., Kelly, M.~J., Lee, M.~E., Rees, P.~K., and Kerr, T.~M., {\it IEEE Electron Device Letters} {\bf 10}, 288 (1989).

\bibitem{Faist1994}
Faist, J.,  Capasso, F., Sivco, D.~L., Sirtori, C., Hutchinson, A.~L. and Cho, A.~Y. Quantum Cascade Laser. {\it Science} {\bf 264}, 553-556 (1994).

\bibitem{Eaves1984}
Eaves, L., Guimaraes, P.~S.~S., Portal, J.~C., Pearsall, T.~P. and Hill, G. High-Field Resonant Magnetotransport Measurements in Small n$^+$nn$^+$ GaAs Structures. {\it Phys. Rev. Lett.} {\bf 53}, 608-611 (1984).

\bibitem{Eaves1984a}
Eaves, L., Guimaraes, P.~S.~S., Portal, J.~C., Hot-electron magnetophonon spectroscopy on micron- and sub-micron-size n$^{+}$nn$^+$ GaAs structures {\it J. Phys. C: Solid State Phys.} {\bf 17}, 6177-6190 (1984).

\bibitem{Mori1988}
Mori, N., Nakamura, N., Taniguchi, K. and Hamaguchi, C. Electric field-induced magnetophonon resonance. {\it Solid. State. Electron.} {\bf 31}, 777-780 (1988).

\bibitem{Zudov2001}
Zudov, M. A. et al. New class of magnetoresistance oscillations: Interaction of a two-dimensional electron gas with leaky interface phonons. {\it Phys. Rev. Lett.} {\bf 86}, 3614-3617 (2001).

\bibitem{Zhang2008}
Zhang, W, Zudov, M.~A., Pfeiffer, L.~N., and West, K.~W. Resonant Phonon Scattering in Quantum Hall Systems Driven by dc Electric Fields. {\it Phys. Rev. Lett.} {\bf 100}, 036805 (2008).

\bibitem{Dmitriev2010}
Dmitriev, I.~A., Gellmann, R. and Vavilov, M.~G.  Phonon-induced resistance oscillations of two-dimensional electron systems drifting with supersonic velocities. {\it Phys. Rev. B} {\bf 82}, 201311(R) (2010).

\bibitem{Dmitriev2012}
Dmitriev, I.~A., Mirlin, A.~D., Polyakov, D.~G. and Zudov, M.~A. Nonequilibrium phenomena in high Landau levels. {\it Rev. Mod. Phys.} {\bf 84}, 1709-1763 (2012).

\bibitem{Yang2002}
Yang, C.~L., Zhang, J., Du, R.~R., Simmons, J.~A. and Reno, J.~L. Zener Tunneling Between Landau Orbits in a High-Mobility Two-Dimensional Electron Gas. {\it Phys. Rev. Lett.} {\bf 89}, 1-4 (2002). 

\bibitem{Zhang2007}
Zhang, W, Chiang, H.~S. , Zudov, M.~A. ,Pfeiffer,  L.~N. and West.
K.~W. \newblock {Magnetotransport in a two-dimensional electron system in dc electric fields}.
\newblock {\em Phys. Rev. B} {\bf 75}, 41304 (2007).

\bibitem{Zudov2017}
Zudov, M.~A., Dmitriev, I.~A., Friess, B., Shi, Q., Umansky, V., {Von Klitzing} K., and Smet. J.
\newblock {Hall field-induced resistance oscillations in a tunable-density GaAs quantum well}.
\newblock {\em Phys. Rev. B} {\bf 96}, 121301 (2017).

\bibitem{Shi2014}
Shi, Q, Ebner, Q.~A. and Zudov, M.~A., Hall field-induced resistance oscillations in a p-type Ge/SiGe quantum well. {\it Phys. Rev. B} {\bf 90} 161301(R) (2014).

\bibitem{Shi2017}
Shi, Q, Zudov, M.~A., Falson, J., Kozuka, Y., Tsukazaki, A., Kawasaki, M., von Klitzing K., and Smet J., Hall field-induced resistance oscillations in MgZnO/ZnO heterostructures. {\it Phys. Rev. B} {\bf 95}, 041411(R) (2017).


\bibitem{Tsui1983} Tsui, D.~C., Dolan, G.~J., and Gossard, A.~C. Zener breakdown of the quantized Hall effect. {\it Bull. Am. Phys. Soc.} {\bf 28}, 365 (1983).

\bibitem{Heinonen1984}
Heinonen, O., Taylor, P.~L., and Girvin, S.~M. Electron-phonon interactions and the breakdown of the dissipationless quantum Hall effect. {\it Phys. Rev. B} {\bf 30}, 3016 (1984).

\bibitem{Eaves1986}
Eaves, L. and Sheard F.~W. Size-dependent quantised breakdown of the dissipationless quantum Hall effect in narrow channels. {\it Semicond. Sci. Technol.} {\bf 1}, 346 (1986).

\bibitem{Tomimatsu2020}
Tomimatsu, T., Hashimoto, K.,  Taninaka, S., Nomura, S. and Hirayama, Y. Probing the breakdown of topological protection: Filling-factor-dependent evolution of robust quantum Hall incompressible phases. {\it Phys. Rev. Res.} {\bf 2}, 013128 (2020).

\bibitem{Martin2003}
Martin, A.~M., Benedict, K.~A., Sheard, F.~W. and Eaves, L. Model for the Voltage Steps in the Breakdown of the Integer Quantum Hall Effect. {\it Phys. Rev. Lett.} {\bf 91}, 126803 (2003).

\bibitem{Novoselov2005} Novoselov, K.~S., Geim, A.~K., Morozov, S.~V., ~Jiang, D.,  Katsnelson, M.~I., 
  Grigorieva, I.~V.,  Dubonos, S.~V., and  Firsov. A.~A.
\newblock {Two-dimensional gas of massless Dirac fermions in graphene}.
\newblock {\em Nature} {\bf 438}, 197-200 (2005).

\bibitem{Zhang2005}
Zhang, Y., Tan, Y.~W., Stormer, H.~L., and Kim, P.
\newblock {Experimental observation of the quantum Hall effect and Berry's
  phase in graphene}.
\newblock {\em Nature} {\bf 438}, 201-204 (2005).

\bibitem{CastroNeto2009}
Castro Neto, A.~H., Guinea, F., Peres, N.~M.~R., Novoselov, K.~S. and Geim, A.~K. The electronic properties of graphene. {\it Rev. Mod. Phys.} {\bf 81}, 109-162 (2009).

\bibitem{Goerbig2011}
Goerbig, M.~O. Electronic properties of graphene in a strong magnetic field. {\it Rev. Mod. Phys.} {\bf 83}, 1193 (2011).

\bibitem{Yankowitz2019}
Yankowitz, M., Ma, Q., Jarillo-Herrero P. and LeRoy, B.~J.  van der Waals heterostructures combining graphene and hexagonal boron nitride. {\it Nat. Rev. Phys.} {\bf 1}, 112 (2019).

\bibitem{KrishnaKumar2017}
Krishna Kumar, R. {\it et al.} High-temperature quantum oscillations caused by recurring Bloch states in graphene superlattices {\it Science} {\bf 357} 181-184 (2017)

\bibitem{Kallin1984}
Kallin, C. and Halperin, B.~I. Excitations from a filled Landau level in the two-dimensional electron gas. {\it Phys. Rev. B} {\bf 30}, 5655-5668 (1984).

\bibitem{Girvin1986}
Girvin, S.~M., MacDonald, A.~H. and Platzman, P.~M. Magneto-roton theory of collective excitations in the fractional quantum Hall effect. {\it Phys. Rev. B} {\bf 33}, 2481-2494 (1986).

\bibitem{Roldan2010}
Rold\'an, R., Goerbig, M. O. and Fuchs, J. N. The magnetic field particle-hole excitation spectrum in doped graphene and in a standard two-dimensional electron gas. {\it Semicond. Sci. Technol.} {\bf 25}, (2010).

\bibitem{Yang2018}
Yang, W. {\it et al.} Landau Velocity for Collective Quantum Hall Breakdown in Bilayer Graphene. {\it Phys. Rev. Lett.} {\bf 121}, 136804 (2018).



\bibitem{Kumaravadivel2019}
Kumaravadivel, P.,  Greenaway, M.~T., Perello, D., Berdyugin, A., Birkbeck, J., Wengraf, J., Liu, S., Edgar, J.~H., Geim, A.~K., Eaves L. and Krishna Kumar, R. Strong magnetophonon oscillations in extra-large graphene.
{\it Nat. Commun.} {\bf 10}, 3334 (2019). 

\bibitem{Greenaway2019}
Greenaway, M.~T., Krishna Kumar, R., Kumaravadivel, P., Geim,  A.~K. and Eaves, L. Magnetophonon spectroscopy of Dirac fermion scattering by transverse and longitudinal acoustic phonons in graphene. {\it Phys. Rev. B} {\bf 100}, 155120 (2019).

\bibitem{Taychatanapat2013}
Taychatanapat, T., Watanabe, K., Taniguchi, T. and Jarillo-Herrero, P.
\newblock {Electrically tunable transverse magnetic focusing in graphene}.
\newblock {\em Nat. Phys} \textbf{9}, 225-229 (2013).

\bibitem{Lukose2007}
Lukose, V., Shankar, R. and Baskaran, G. Novel Electric Field Effects on Landau Levels in Graphene. {\it Phys. Rev. Lett.} {\bf 98}, 116802 (2007).

\bibitem{Raichev2020}
Raichev, O.~E. and Zudov, M.~A. Effect of Berry phase on nonlinear response of two-dimensional fermions. {\it Phys. Rev. Res.} {\bf 2} 022011(R) (2020).

\bibitem{Greenaway2015}
Greenaway, M.~T. {\it et al.} Resonant tunnelling between the chiral Landau states of twisted graphene lattices. {\it Nat. Phys.} {\bf 11}, 1057-1062 (2015).

\bibitem{Esaki1962}
Esaki, L. New Phenomenon in Magnetoresistance of Bismuth at Low Temperature {\it Phys. Rev. Lett.} {\bf 8} 4 (1962).

\bibitem{Hutson1961}
Hutson, A.~R., McFee, J.~H. and White, D.~L. Ultrasonic amplification in CdS {\it Phys. Rev. Lett.} {\bf 7} 237 (1961).

\bibitem{Spector1963}
Spector, H.~N. Magnetic Field Dependence of the Amplification of Sound by Conduction Electrons {\it Phys. Rev.} {\bf 131}, 2512 (1963).

\bibitem{Eckstein1963}
Eckstein, S.~G. Resonant Amplification of Sound by Conduction Electrons {\it Phys. Rev.} {\bf 131} 1087 (1963).

\bibitem{Funk2015}
Funk, H., Knorr, A., Wendler, F. and Malic, E. Microscopic view on Landau level broadening mechanisms in graphene. {\it Phys. Rev. B} {\bf 92}, 205428 (2015).

\bibitem{Zener1934}
Zener, C. A theory of the electrical breakdown of solid dielectrics. {\it Proc. R. Soc. Lond. A} {\bf 145}, 523-529 (1934). 

\bibitem{Vavilov2007}
Vavilov, M.~G., Aleiner, I.~L. and Glazman, L.~I.  Nonlinear resistivity of a two-dimensional electron gas in a magnetic field {\it Phys. Rev. B} {\bf 76} 115331 (2007)

\bibitem{Landau1941}
Landau, L. Theory of the Superfluidity of Helium II. {\it Phys. Rev.} {\bf 60}, 356-358 (1941).

\bibitem{Trachenko2020}
Trachenko, K. and Brazhkin V.~V. Minimal quantum viscosity from fundamental physical constants. {\it Sci. Adv.} {\bf 6}, eaba3747 (2020).

\bibitem{Lvov2014}
L'vov V.~S., Skrbek L., and Sreenivasan K.~R. Viscosity of liquid 4He and quantum of circulation: Are they related? {\it Physics of Fluids} {\bf 26}, 041703 (2014).

\end{thebibliography}
\end{document}